\documentclass[11pt]{article}
\usepackage{graphics,amsmath,bbm,url}

\newcommand{\im}{{\rm i}}
\newcommand{\C}{\mathbbm{C}}
\newcommand{\N}{\mathbbm{N}}
\newcommand{\dif}{{\rm d}}
\newcommand{\Prob}{\mathop{\rm\bf\null P}\nolimits}
\makeatletter
\newcommand{\address}[1]{\long\edef\@title{\@title\\{\it #1}}}
\makeatother

\title{Monte Carlo study of the hull distribution for the $q=1$ Brauer model}
\author{Wouter Kager\footnote{Present address: EURANDOM, P.O. Box 513,
5600 MB Eindhoven, the Netherlands}\ \ and Bernard Nienhuis\\\null\\
\it Instituut voor Theoretische Fysica\\
\it Valckenierstraat 65\\
\it 1018 XE Amsterdam, the Netherlands\\\null\\
kager@eurandom.tue.nl, nienhuis@science.uva.nl}
\date{}

\begin{document}

\maketitle

\begin{abstract}
We study a special case of the Brauer model in which every path of the model
has weight~$q=1$. The model has been studied before as a solvable lattice
model and can be viewed as a Lorentz lattice gas. The paths of the model are
also called self-avoiding trails. We consider the model in a triangle with
boundary conditions such that one of the trails must cross the triangle from
a corner to the opposite side. Motivated by similarities between this model,
SLE(6) and critical percolation, we investigate the distribution of the hull
generated by this trail (the set of points on or surrounded by the trail) up
to the hitting time of the side of the triangle opposite the starting point.
Our Monte Carlo results are consistent with the hypothesis that for system
size tending to infinity, the hull distribution is the same as that of a
Brownian motion with perpendicular reflection  on the boundary.
\end{abstract}

\section{Introduction to the model}
\label{sec:brauer}

In this paper we present results from a Monte Carlo study of a special case
of the Brauer model. This model has appeared in the literature in different
guises, and received the name Brauer model only recently. Originally, it was
studied as a $q$-state solvable vertex model~\cite{schultz:1981} and later
as an O($q$) symmetric, solvable lattice model~\cite{reshetikhin:1983,
devega:1987}. More recently it was observed that the model could be seen as
a model of paths on the lattice in which each closed path has a weight equal
to~$q$, where~$q$ can take non-integer values. In this language of paths
each vertex can carry one of the following configurations of path segments,
with the corresponding weights:
\begin{center}
\begin{picture}(220,30)
\put(0,10){\includegraphics{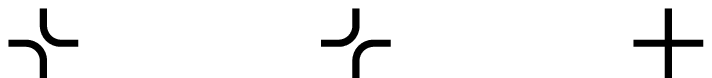}}
\put(20,0){\makebox(0,0){$u$}}
\put(110,0){\makebox(0,0){$1-u$}}
\put(200,0){\makebox(0,0){$(1-q/2)u(1-u)$}}
\end{picture}
\end{center}
Here the weights are chosen to be solutions of the Yang-Baxter (or
star-triangle) equation~\cite{baxter:1978,baxter:1982}. Although the
integrability condition in the Yang-Baxter equation does not restrict~$q$,
the actual solution by means of the Bethe Ansatz was only extended to the
integers~\cite{martins:1998}.

The name Brauer model was given to emphasise that the transfer matrix of
the model is an element of the Brauer algebra~\cite{degier:2005}. The model
attracted particular attention in the limit $q\to1$ when it describes the
probability distribution of lattice paths taken by a particle that is
scattered by randomly placed, static scatterers~\cite{gunn:1985,kong:1989,
ruijgrok:1988,ziff:1991}. As such it can be viewed as a Lorentz lattice
gas, although in these applications the vertex weights are usually not
chosen as above.

Another paradigm is coming from the analogy to the self-avoiding walk, as
in the Brauer model a walk is not permitted to traverse a lattice edge more
than once. A walk subject to this condition and with no further restrictions
on the vertices is called a self-avoiding trail (SAT)~\cite{lyklema:1985}. We
will therefore refer to the paths in the Brauer model as \emph{trails}. Here
we study the model at $q=1$ on a bounded domain, and we will be interested in
the distributions of special points on the boundary that are visited by one of
the trails. This will be explained in more detail below.

\begin{figure}
\centering
\begin{picture}(180,184)
\put(0,0){\includegraphics{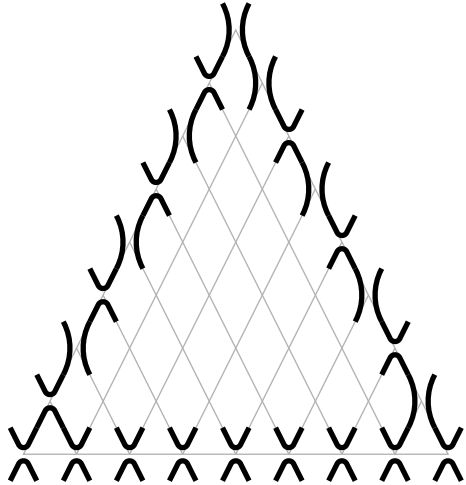}}
\end{picture}
\caption{Boundary conditions for the self-avoiding trail in a triangle.}
\label{fig:BrauerTriangle}
\end{figure}

First we introduce some notation. For given angles $\alpha$ and~$\beta$ in
the range $(0,\pi)$ such that $\alpha+\beta<\pi$, we define $T_{\alpha,\beta}$
as the triangle in the upper half of the complex plane with vertices at $0$
and~$1$, such that the interior angle at~$0$ is~$\alpha$ and the interior
angle at~$1$ is~$\beta$. By $w_{\alpha,\beta}$ we will denote the third
vertex of~$T_{\alpha,\beta}$. For a given angle $\phi\in(0,\pi/2)$ and
integer system size $N>0$, let $V=V_\phi$ be the set of vertices $\{2j\cos
\phi+k\exp(i\phi):j,k\in\N,j+k\leq N\}$. Properly rescaled, this collection
of vertices provides a nice covering of the isosceles triangle
$T_{\phi,\phi}$ with base angle~$\phi$, see figure~\ref{fig:BrauerTriangle}.

We define the Brauer model on~$V$ with $q=1$ as follows. Each vertex of~$V$
can carry either of the three following configurations of trail segments:
\begin{center}
\begin{picture}(220,22)
\put(0,0){\includegraphics{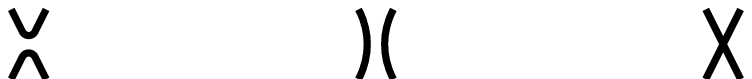}}
\put(-18,8){$a=\null$}
\put(83,8){$b=\null$}
\put(183,8){$c=\null$}
\label{vertexstates}
\end{picture}
\end{center}
Here, the third vertex state should be interpreted as a crossing. The state of
each vertex is chosen from $\{a,b,c\}$ independently from the states of the
other vertices, with the probabilities for the three states given by
\begin{equation}
 p(a) = \lambda\pi\,2\phi \qquad
 p(b) = \lambda\pi\,(\pi-2\phi) \qquad
 p(c) = \lambda\,\phi(\pi-2\phi)
 \label{equ:weights}
\end{equation}
where $\lambda:=[\pi^2+\phi(\pi-2\phi)]^{-1}$ provides the normalisation.
Two distant edges may or may not be connected to each other by a trail. The
correlations of these events are expected to be isotropic in space for large
distances, if the weights are chosen as in equation~(\ref{equ:weights}) and
vertices are arranged in space as in figure~\ref{fig:BrauerTriangle}. The
anisotropy of the weights is thus precisely compatible with the spatial
anisotropy of the lattice, see~\cite{kim:1987}.

Each configuration of vertex states defines a collection of self-avoiding
trails on the vertex set~$V$. We will be interested only in those
configurations in which one of these trails crosses the triangular domain
from a given corner to the opposite side. To make this trail stay inside
the triangle we have to impose suitable boundary conditions. We choose the
boundary conditions as shown in figure~\ref{fig:BrauerTriangle}. Here it is
assumed that the system size~$N$ is an even number, so that each side of the
triangle carries an odd number of vertices. With these boundary conditions,
one trail of the model must enter the triangle at the top or the lower-left
corner, and stay inside the triangle until it leaves at the other of these
two corners.

In our simulations we generate this particular self-avoiding trail
dynamically as follows. Initially, only the vertex states on the boundary are
fixed according to the boundary conditions. The vertex states in the interior
are still undecided. The trail starts from either the top or the lower-left
corner of the triangle. The steps of the trail follow the trail segments of
the vertex states. Each time the trail meets a vertex whose state is still
undetermined, we choose its state according to the probabilities $p(a)$,
$p(b)$ and~$p(c)$ given above, and continue the random walk. The state of
this vertex is fixed forever onwards. We stop the simulation as soon as the
trail hits a vertex on the side of the triangle opposite the starting point.

We want to study the distribution of the point where the trail hits the side
opposite the starting point in the limit when~$N$ becomes large. In fact, we
are interested in the distribution of the so-called \emph{hull} generated by
the trail up to the stopping time, that is, the collection of points in the
triangle that are disconnected by the stopped trail from the side of the
triangle opposite the starting point. This will be motivated in more detail
in section~\ref{sec:locality}. As we shall explain, the main motivation for
our study of the hull distribution is the close connection between the
Brauer model and critical bond percolation, for which the hull distribution
is known (although rigorously only for critical site percolation on the
triangular lattice).

To make the connection with bond percolation, let us modify the model above
by setting $p(c)=0$, and let $p(a)$ and~$p(b)$ be proportional to $\sin(2
\phi/3)$ and $\sin((\pi-2\phi)/3)$, respectively (see for instance
\cite[section~5]{kenyon:2003}). For any given configuration of the model,
one can draw at every vertex either the horizontal or the vertical edge
connecting the centres of two adjacent rhombi (see
figure~\ref{fig:BrauerTriangle}), such that the drawn edge does not intersect
a trail segment at that vertex. It is easy to see that the drawn edges
constitute a configuration of bond percolation on a rectangular lattice
together with the dual configuration. On the boundary we have a percolation
cluster along the base and right side of the triangle and a dual cluster
along the left side. Note that as before, there is one special path in the
model that crosses the triangle between the top and the lower-left corner,
and that this path describes the interface between the cluster and the dual
cluster attached to the boundary. This path is called the percolation
exploration process.

Thus we can interpret the Brauer model at $q=1$ as a variant of a loop model
for percolation with the added possibility that the loops may cross. Note
that these crossings happen only with small probability, since $p(c)$ is the
smallest of the three weights. One can therefore expect a self-avoiding trail
to explore space much like the exploration process of critical percolation,
except that occasionally the trail may cross and possibly re-enter a part of
space it has already explored.

\section{Conformal invariance and locality}
\label{sec:locality}

In section~\ref{sec:brauer} we introduced the Brauer model. It is believed
that the scaling limit of this model is conformally invariant. Moreover, a
self-avoiding trail of the model has the locality property. The purpose of
this section is to explain what we precisely mean by these two properties,
and to discuss an important implication.

To explain conformal invariance, suppose that for every domain in~$\C$ with
continuous boundary and suitable boundary conditions, we are given a
probability measure on the collection of paths in that domain. Then we say
that this family of measures (or the family of random paths they describe)
is conformally invariant if for any conformal map~$g$ from a domain~$D$ onto
a domain~$D'$ that maps corresponding boundary conditions onto each other,
the image of the measure on paths in~$D$ by~$g$ coincides with the measure
on paths in~$D'$.

\begin{figure}
\includegraphics{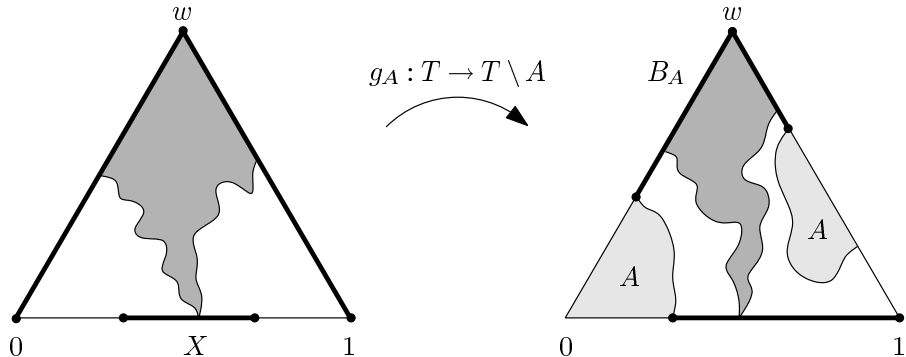}
\caption{A hull generated by a local path in a triangle~$T$. The
 distribution of the hull is determined by the probabilities that it avoids
 sets~$A$ such as shown on the right.}
\label{fig:LocalityHull}
\end{figure}

Now let~$T$ be a triangle in the upper half of the complex plane with vertices
at $0$, $1$ and~$w$, and suppose that~$D$ is of the form $T\setminus A$,
where~$A\subset T$ is such that $T\setminus A$ is simply connected and
$w\in(\partial T\setminus\partial A)$, see figure~\ref{fig:LocalityHull}. Let
the map $g_A:T\to T\setminus A$ fix~$w$ and send $\partial T\setminus[0,1]$
onto the boundary arc~$B_A$ of $\partial T\setminus\partial A$ containing~$w$.
Consider a random path~$W$ in this triangle which starts in~$w$ and is
stopped as soon as it hits the interval $[0,1]$. Then we say that~$W$ has
the locality property if the path in~$T$ started from~$w$ and stopped at the
first time when it hits $\partial D\setminus B_A$, has the same distribution
as the path in~$D$ started from~$w$ and stopped on $\partial D\setminus B_A$.
Note that if~$W$ is conformally invariant, then the latter random path is the
image of the former by~$g_A$.

For a local, conformally invariant path~$W$ in~$T$, let~$\tau$ denote the
first time when~$W$ hits the interval~$[0,1]$, and set $X:=W(\tau)$. Then we
define the \emph{hull}~$K$ generated by the path as the set of points
in~$\overline{T}$ that are either on $W[0,\tau]$ or are disconnected from
$\{0,1\}$ by $W[0,\tau]$. Observe that the distribution of this hull~$K$ is
determined if we know for all sets~$A$ as in the previous paragraph the
probability that $K\cap A=\emptyset$. But by the locality property, this
probability is exactly the probability that the path~$W$ in~$D=T\setminus
A$ stopped when it hits $\partial D\setminus B_A$, is stopped in the
interval~$[0,1]$. Using conformal invariance, this probability equals the
probability that~$X$ is in the interval~$g_A^{-1}([0,1]\setminus\partial A)$,
see figure~\ref{fig:LocalityHull}. It follows that for local paths, the
distribution of the hull~$K$ is determined by the distribution of the exit
point~$X$ (see~\cite{werner:2001} for an illuminating discussion).

To examine some concrete examples, let~$T$ be an equilateral triangle (in
other words, take $w=\exp(\im\pi/3)$). Then there are three known local and
conformally invariant paths such that~$X$ has the uniform distribution. The
first example is the trace of Schramm-L\"owner Evolution (SLE) for $\kappa=6$
\cite{lsw:2003,werner:2004b}. The second example is a Brownian motion in~$T$
which is reflected on the left side in the direction given by the vector
$\exp(-\im\pi/3)$ and on the right side in the direction given by
$\exp(-2\im\pi/3)$~\cite{lsw:2001a,werner:2004b} (for details on reflected
Brownian motion, see section~\ref{sec:rbm}). The third example is the scaling
limit of the exploration process for critical site percolation on the
triangular lattice in~$T$~\cite{smirnov:2001}. Since the distribution of~$X$
is the same for these paths, by the result of the previous paragraph they
generate hulls~$K$ with the same distribution.

As we explained in section~\ref{sec:brauer}, the Brauer model studied here
is closely related to critical percolation. One may therefore expect the
distribution of the hull to be the same for the two models. This can be
motivated further by the fact that any two local and conformally invariant
processes in the plane which are started in the origin and stopped upon
hitting the unit circle generate the same hull, namely that of a stopped
Brownian motion (see~\cite{werner:2001}). In a bounded domain, however, our
Monte Carlo study shows that the hull distributions for critical percolation
and for the Brauer model are not the same. Instead, one of the results of
this paper is that our numerical data are consistent with the hypothesis that
in the scaling limit, the hull of the Brauer model has the same distribution
as that of a Brownian motion which is reflected perpendicularly on the
boundary of the domain.

\begin{figure}
\centering\includegraphics{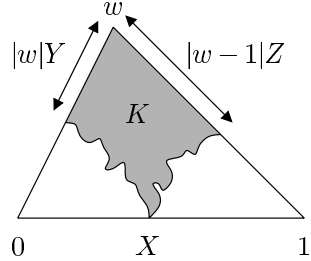}
\caption{Definition of the hull~$K$ of a local path in the triangle~$T$,
 and of the random variables $X$, $Y$ and~$Z$.}
\label{fig:LocalXYZ}
\end{figure}

To study the distribution of the hull for the Brauer model, by locality it
is in principle sufficient to look only at the distribution of the exit
point~$X$ in a given triangle. However, for obvious reasons we prefer to
consider more characteristics of the hull. For this purpose we introduce
two new random variables $Y$ and~$Z$ associated with the hull~$K$ in the
triangle~$T$, as follows. We shall denote by $|w|Y$ the distance of the
lowest point of the hull on the left side of~$T$ to the top~$w$, and by
$|w-1|Z$ the distance of the lowest point of the hull on the right side to
the top~$w$. Thus, all three random variables $X$, $Y$ and~$Z$ take values
in the range~$[0,1]$. See figure~\ref{fig:LocalXYZ} for an illustration.

In section~\ref{sec:rbm} we shall compute the (joint) distributions of $X$,
$Y$ and~$Z$ for the case of reflected Brownian motion. These distributions
can then be compared with the corresponding distributions in the Brauer
model. The results of our Monte Carlo study of the hull distribution are
discussed in section~\ref{sec:brauerdist}. In section~\ref{sec:lastvisit}
we present simulation results for a different distribution, the last-visit
distribution.

\section{Reflected Brownian motion}
\label{sec:rbm}

In this section we define reflected Brownian motion in triangles and
summarize the main results on these processes. We recall that for given
angles $\lambda$ and~$\mu$ in the range $(0,\pi)$ such that $\lambda+\mu
<\pi$, we define $T=T_{\lambda,\mu}$ as the triangle in the upper half of
the complex plane with vertices at $0$ and~$1$, such that the interior
angle at~$0$ is~$\lambda$ and the interior angle at~$1$ is~$\mu$. By $w=
w_{\lambda,\mu}$ we denote the third vertex of~$T_{\lambda,\mu}$.

Let us now introduce the (reflection) angles $\alpha,\beta\in(0,\pi)$ and
set $v_L:=\exp(\im(\lambda+\alpha-\pi))$, $v_R:=\exp(-\im(\mu+\beta))$. We
want to consider a stochastic process~$Z$ in~$T$ which is Brownian motion
in the interior, and which is reflected instantaneously in the direction
given by $v_L$ or~$v_R$ when it hits the left or right side of~$T$,
respectively. Such a process is called a reflected Brownian motion with
reflection vector fields $v_L$ and~$v_R$ on the sides of~$T$. We write
RBM$_{\alpha,\beta}$ to denote this process. Note that the angles $\alpha$
and~$\beta$ are the reflection angles measured with respect to the boundary.

More explicitly, an RBM$_{\alpha,\beta}$ in~$T$ may be defined as follows.
Let~$B$ be standard two-dimensional Brownian motion. Then an RBM$_{\alpha,
\beta}$ in~$T$ is the unique process~$Z$ such that
\begin{equation}
 Z(t)=B(t)+v_L\,Y_L(t)+v_R\,Y_R(t),
\end{equation}
where $Y_L,Y_R$ are continuous increasing real-valued processes adapted
to~$B$ which satisfy $Y_L(0)=Y_R(0)=0$ and increase only when~$Z$ is on the
left, respectively right, side of~$T$ \cite[equation~(2.4)]{varadhan:1985}.
This process is well-defined up to the first time when~$Z$ hits the
interval~$[0,1]$. For a characterisation and properties of these reflected
Brownian motions, see Varadhan and Williams~\cite{varadhan:1985}. It is
straightforward to show conformal invariance and locality for these
processes, see for instance~\cite[chapter~5]{werner:2004b}.

It was shown by Lawler et al.~\cite{lsw:2003} that if we take an
RBM$_{\pi/3,\pi/3}$ in the equilateral triangle~$T_{\pi/3,\pi/3}$, then the
exit distribution is uniform. Their arguments were generalised to isosceles
triangles (i.e.\ the triangles $T_{\lambda,\lambda}$) by
Dub\'edat~\cite{dubedat:2004a}, and then to the case of a generic triangle
(any $T_{\lambda,\mu}$ such that $\lambda+\mu<\pi$, with a natural extension
to the case $\lambda+\mu\geq\pi$) by one of us~\cite{kager:2004}. These
results show that an RBM$_{\alpha,\beta}$ in the triangle~$T_{\alpha,\beta}$
has the uniform exit distribution. By conformal invariance, this also
determines the exit distribution for an RBM$_{\alpha,\beta}$ in any
triangle~$T_{\lambda,\mu}$, and using locality we can in fact compute the
joint distribution of the exit point~$X$ and the lowest points $Y$ and~$Z$
of the hull on the two sides, as we show next.

It turns out that the distributions of the random variables $X$, $Y$ and~$Z$
can all be expressed in terms of conformal transformations of the upper
half-plane onto triangles $T_{\gamma,\delta}$. By the Schwarz-Christoffel
formula~\cite{ahlfors:1966}, the unique conformal transformation of the
upper half-plane onto~$T_{\gamma,\delta}$ that fixes $0$ and~$1$ and
maps~$\infty$ to~$w_{\gamma,\delta}$ is given by
\begin{equation}
 F_{\gamma,\delta}(z) =
  \frac{\int_0^z t^{\gamma/\pi-1}(1-t)^{\delta/\pi-1}\dif{t}}
   {\int_0^1 t^{\gamma/\pi-1}(1-t)^{\delta/\pi-1}\dif{t}}\,.
 \label{equ:trianglemapping}
\end{equation}
By the substitution $t\mapsto1-u$ it is easy to show that
\begin{equation}
 F_{\gamma,\delta}(z) = 1-F_{\delta,\gamma}(1-z) \quad\mbox{and}\quad
 F^{-1}_{\gamma,\delta}(z) = 1-F^{-1}_{\delta,\gamma}(1-z).
 \label{equ:symmetry}
\end{equation}
We shall now explain how we can express the distribution functions for $X$,
$Y$ and~$Z$ in terms of these conformal transformations.

\begin{figure}
\includegraphics{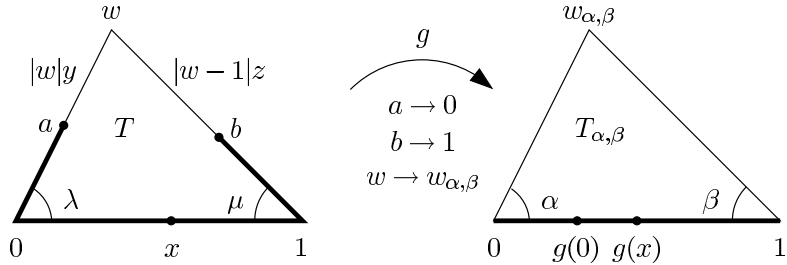}
\caption{This figure illustrates how the joint distribution function of the
 random variables $X$, $Y$ and~$Z$ can be computed. As explained in the text,
 the joint probability $\Prob[X\leq x,Y\leq y,Z\leq z]$ is just $g(x)-g(0)$.}
\label{fig:JointSmall}
\end{figure}

The idea of the computation of $\Prob[X\leq x,Y\leq y,Z\leq z]$ is
illustrated in figure~\ref{fig:JointSmall}. Consider an RBM$_{\alpha,\beta}$
in the triangle~$T=T_{\lambda,\mu}$ started from the top~$w=w_{\lambda,\mu}$.
Let $a=a(y)$ and~$b=b(z)$ be the points on the left and right sides of~$T$
at distances $|w|y$ and~$|w-1|z$ from~$w$, respectively. Stop the RBM as
soon as it hits the counter-clockwise arc from $a$ to~$b$ on the boundary
(the thick line in figure~\ref{fig:JointSmall}). Then the probability
$\Prob[X\leq x,Y\leq y,Z\leq z]$ is just the probability that this process
is stopped in the interval~$(0,x)$.

\begin{figure}
\centering\includegraphics{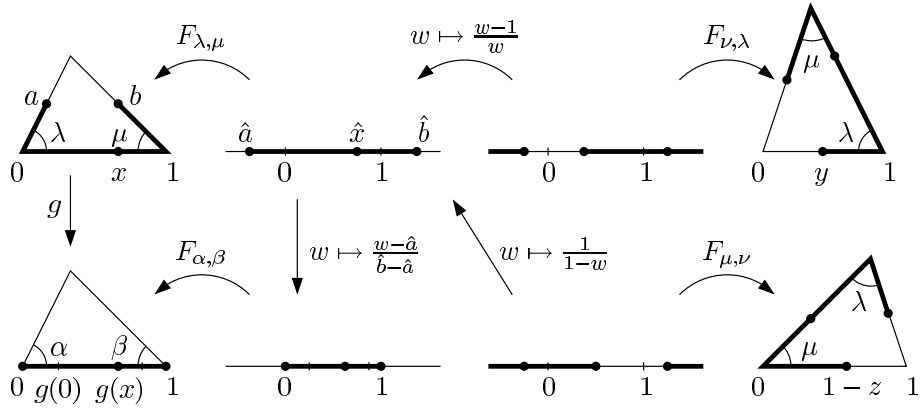}
\caption{This illustration shows schematically how one obtains an explicit
 form for the map~$g$ in terms of the variables $y$ and~$z$. The notations
 $\hat{a}$, $\hat{b}$ and~$\hat{x}$ in the figure are short for
 $F^{-1}_{\lambda,\mu}(a)$, $F^{-1}_{\lambda,\mu}(b)$
 and~$F^{-1}_{\lambda,\mu}(x)$.}
\label{fig:JointBig}
\end{figure}

We now use conformal invariance and locality. Let $g=g_{a(y),b(z)}$ be the
conformal map of~$T$ onto~$T_{\alpha,\beta}$ that sends $a$ to~$0$, $b$
to~$1$ and $w$ to~$w_{\alpha,\beta}$, as illustrated in
figure~\ref{fig:JointSmall}. Then we are looking for the probability that an
RBM$_{\alpha,\beta}$ in~$T_{\alpha,\beta}$ started from~$w_{\alpha,\beta}$
and stopped when it hits~$[0,1]$, is stopped in the interval~$(g(0),g(x))$.
But since the exit distribution of the RBM$_{\alpha,\beta}$ is uniform
in~$T_{\alpha,\beta}$, this probability is simply $g(x)-g(0)$. Thus,
\begin{equation}
 \Prob[X\leq x,Y\leq y,Z\leq z] = g(x)-g(0).
\end{equation}
It remains to find an explicit form of the map~$g=g_{a(y),b(z)}$. At this
point it is useful to denote by~$\nu$ the third angle of the triangle
$T_{\lambda,\mu}$, that is, $\nu:=\pi-\lambda-\mu$. The explicit form
of~$g$ can be obtained by suitably combining conformal self-maps of the
upper half-plane with triangle mappings. How this is done exactly is
described in figure~\ref{fig:JointBig}.

By studying this figure we obtain the formula
\begin{multline}
 \Prob[X\leq x,Y\leq y,Z\leq z]=\null\\
 F_{\alpha,\beta}\left(
     \frac{F^{-1}_{\lambda,\mu}(x)-F^{-1}_{\lambda,\mu}(a)}
	      {F^{-1}_{\lambda,\mu}(b)-F^{-1}_{\lambda,\mu}(a)}
     \right)
 -F_{\alpha,\beta}\left(
     \frac{-F^{-1}_{\lambda,\mu}(a)}
	      {F^{-1}_{\lambda,\mu}(b)-F^{-1}_{\lambda,\mu}(a)}
     \right)
 \label{equ:joint}
\end{multline}
where the images of $a$ and~$b$ under $F^{-1}_{\lambda,\mu}$ can be
expressed in terms of $y$ and~$z$ as
\begin{equation}
 F^{-1}_{\lambda,\mu}(a)
  = 1 - \frac{1}{F^{-1}_{\nu,\lambda}(y)}\,; \qquad
 F^{-1}_{\lambda,\mu}(b)
  = \frac{1}{1-F^{-1}_{\mu,\nu}(1-z)}
  = \frac{1}{F^{-1}_{\nu,\mu}(z)}\,.
\end{equation}
Sending two of the variables $x,y,z$ to~$1$ and using the symmetry
property~(\ref{equ:symmetry}) of the triangle mappings, we obtain
\begin{align}
 & \Prob[X\leq x] = F_{\alpha,\beta}\big( F^{-1}_{\lambda,\mu}(x) \big);\\
 & \Prob[Y\leq y] = F_{\beta,\alpha}\big( F^{-1}_{\nu,\lambda}(y) \big);\\
 & \Prob[Z\leq z] = F_{\alpha,\beta}\big( F^{-1}_{\nu,\mu}(z) \big).
\end{align}
Observe that these three distributions have particularly simple forms with a
nice geometric interpretation. For instance, $\Prob[Y\leq y]$ is just the
image of~$y$ under the transformation that maps the triangle $T_{\nu,\lambda}$
onto~$T_{\beta,\alpha}$, fixes $0$ and~$1$ and takes $w_{\nu,\lambda}$
onto~$w_{\beta,\alpha}$.

\section{Hull distribution of the Brauer model}
\label{sec:brauerdist}

We now return to the Brauer model introduced in section~\ref{sec:brauer}. To
compare the hull of the model with the hull of reflected Brownian motion,
we measure the (joint) distributions of the random variables $X$, $Y$ and~$Z$
(introduced in section~\ref{sec:locality}) in our simulations of the Brauer
model. The data are collected for $10^6$ independent trails starting from the
top and $10^6$ independent trails starting from the lower-left corner on the
set of vertices~$V=V_\phi$ for the 12~different system sizes~$N=\null$100,
200, 300, 400, 600, 800, 1200, 1600, 2000, 2400, 3200, 4000 and 8~different
base angles~$\phi$ ranging from $10^\circ$ up to $80^\circ$ with
$10^\circ$~intervals. Below we shall describe more precisely how the data for
the (joint) distributions of $X$, $Y$ and~$Z$ are collected.

\begin{figure}
\centering\includegraphics{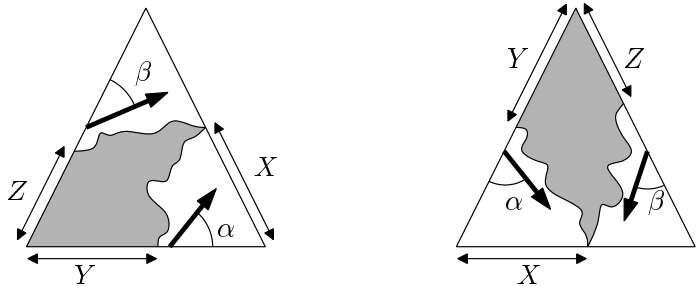}
\caption{How the random variables $X$, $Y$ and~$Z$ and reflection angles
  $\alpha$ and~$\beta$ are associated to the trails starting from different
  corners. For simplicity we have omitted the normalisation of $X$, $Y$
  and~$Z$.}
 \label{fig:BrauerXYZ}
\end{figure}

We distribute the vertices of~$V$ on the three sides of the triangle over
a total of $100$~bins, each containing $N/100$ vertices. Since there are
actually $N+1$ vertices on each side, this means that one vertex on each
side is omitted. This is the vertex in the corner of the side where the
associated random variable $X$, $Y$ or~$Z$ is~$0$. In the simulations we
record the number of trails~$X_i$ that land in the $i$th bin for each
$i=1,2,\ldots,100$. We also record the numbers of walks $Y_i$, respectively
$Z_i$, such that the $i$th bin on the right, respectively left, side of the
triangle (as seen from the starting point) is the bin furthest from the
starting point which was visited by the trail. Figure~\ref{fig:BrauerXYZ}
illustrates how the variables $X$, $Y$ and~$Z$ are associated with the hulls
of trails starting from different corners. The joint distributions of these
variables are recorded similarly, but instead of using $100$~bins on the
sides we use $50$~bins to reduce memory requirements.

In this way the simulations build up histograms of the marginal and joint
distributions of $X$, $Y$ and~$Z$ for the self-avoiding trail that can be
compared to the corresponding distributions for the reflected Brownian
motions. To improve the statistics, we first merge together bins of
collected data in order to distribute the numbers of trails in different
bins more evenly. We shall explain below how this merging procedure works for
the joint distribution of $X$, $Y$ and~$Z$. A similar procedure is applied
for the joint distributions of two of the three random variables and for the
marginal distributions.

For the joint distribution of $X$, $Y$ and~$Z$, from our simulations we have
a total of $50\times50\times50$ cubical bins that span the unit cube with
the variables $X$, $Y$ and~$Z$ along the axes. We consider merging together
either $10\times10\times10$, or $5\times5\times5$, or $2\times2\times2$ of
these cubical bins into larger cubical bins, that together with the unmerged
bins form a partition of the unit cube. Our aim is that each bin in the final
partition of the unit cube represents at least $100$ generated trails,
$0.01\%$ of the total. Moreover, we want each bin that represents at least
$1\%$ ($10^4$) of the total number of generated trails to be present in the
final partition.

The merging procedure therefore works as follows. First, we consider for
each of the $10\times10\times10$ cubes whether it should form a large bin,
or be built up from smaller bins. So we do the following test: if we would
built it up from $2\times2\times2$ bins, then would each constituting bin
contain at least $100$ trails, or would one of these bins contain at least
$10^4$ trails? If so, the $10\times10\times10$ cube is split into
$2\times2\times2$ cubes. Then we test whether each of these $2\times2\times2$
cubes should form a bin, or be built up from $1\times1\times1$ bins, by the
same criterion as above. Otherwise, we test whether the $10\times10\times10$
cube can be built up from $5\times5\times5$ (rather than $2\times2\times2$)
cubes. If so, we test whether each of these $5\times5\times5$ cubes should
form a bin, or be built up from $1\times1\times1$ bins, as before. In the
end, we achieve a more even distribution of trails over a smaller number of
bins, as desired.

\begin{figure}
\centering\includegraphics{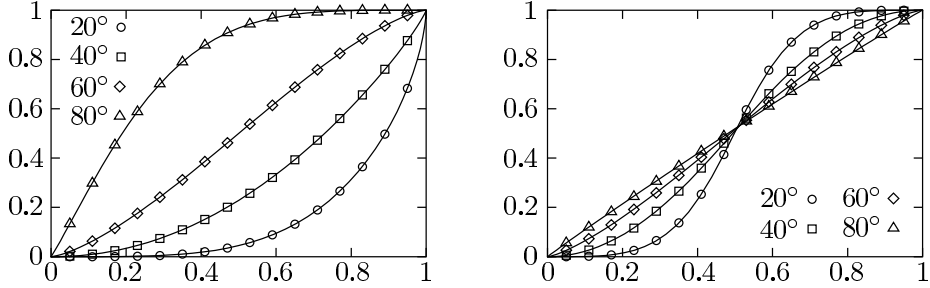}
\caption{The distribution functions $\Prob[X\leq x]$ (left) and $\Prob[
 Y\leq y]$ (right) for the self-avoiding trail on different triangles with
 system size $N=4000$,  started from the lower-left corner. The solid lines
 are the corresponding distributions for a reflected Brownian motion for
 which we obtain an optimal least-squares fit. See
 table~\ref{tab:FittedValues} for the corresponding values of $\alpha$
 and~$\beta$.}
\label{fig:Compare}
\end{figure}

As we explained, we want to compare the distribution of the hull generated by
the trails with the hull distribution of reflected Brownian motion, in
particular in the limit when the system size becomes large. Our working
hypothesis is therefore that the binned data collected in the simulations is
predicted by the joint distribution function~(\ref{equ:joint}) for the RBM
with parameters $\alpha$, $\beta$ for some $\alpha$ and~$\beta$ as $N\to
\infty$. To test this hypothesis, we make a least-squares fit of the
parameters $\alpha$ and~$\beta$ for each system size~$N$. For every triangle
$T_{\phi,\phi}$ on which we have simulated the model and for each of the two
possible starting points, this gives us a list of the best-fit values for
$\alpha$ and~$\beta$ for 12~different system sizes. We want to investigate
whether these values of $\alpha$ and~$\beta$ converge to a common limit as
the system size becomes larger and larger.

The least-squares fits show that the marginal distributions of $X$, $Y$
and~$Z$ for the self-avoiding trail are well described by those of an RBM
(see figure~\ref{fig:Compare}). More precisely, as the system size~$N$ gets
larger, the value of $\chi^2$ (the difference between the actual and the
predicted number of walks in each bin squared, divided by the predicted
variance in this number, summed over all the bins) goes down to a value close
to the number of bins, see for instance table~\ref{tab:Equilateral}. However,
the fitted values we obtain for $\alpha$ and~$\beta$ are not constant with
the system size. More work is therefore needed to test convergence of
$\alpha$ and~$\beta$ to a limit as $N\to\infty$.

We also observe from the simulations that the fitted values of $\alpha$
and~$\beta$ for the three distributions of $X$, $Y$ and~$Z$ do not fully
agree. It seems that the fitted values of~$\alpha$ do agree for the
distributions of $X$ and~$Y$, whereas the fitted values of~$\beta$ do agree
for the distributions of $X$ and~$Z$. The fitted value of~$\alpha$ for the
distribution of~$Z$ agrees with the fitted value of~$\beta$ for the
distribution of~$Y$. See table~\ref{tab:FittedValues}. This observation holds
both for trails starting from the top of the triangle and for trails starting
from the lower-left corner at different system sizes. This is another
indication that more work is required to test convergence of $\alpha$
and~$\beta$ as $N\to\infty$.

\begin{table}
 \centering\small
 \begin{tabular}
  {@{}l|c@{\ }c|c@{\ }c|c@{\ }c|c@{\ }c@{}}
   & \multicolumn{2}{c|}{$20^\circ$} & \multicolumn{2}{c|}{$40^\circ$}
   & \multicolumn{2}{c|}{$60^\circ$} & \multicolumn{2}{c} {$80^\circ$} \\
   & $\alpha/\pi$ & $\beta/\pi$ & $\alpha/\pi$ & $\beta/\pi$
   & $\alpha/\pi$ & $\beta/\pi$ & $\alpha/\pi$ & $\beta/\pi$ \\
  \hline
  \vrule height2.5ex width0pt
  $X$
   & 0.456(1) & 0.484(2) & 0.461(2) & 0.474(2)
   & 0.464(2) & 0.464(2) & 0.464(2) & 0.453(2) \\
  $Y$
   & 0.456(1) & 0.446(1) & 0.462(2) & 0.445(1)
   & 0.464(1) & 0.438(2) & 0.461(2) & 0.431(2) \\
  $Z$
   & 0.447(2) & 0.483(2) & 0.445(1) & 0.475(2)
   & 0.439(2) & 0.464(2) & 0.431(1) & 0.452(1) \\
 \end{tabular}
 \caption{Fitted values for $\alpha$ and~$\beta$ on different triangles
  compared for the different marginal distributions of $X$, $Y$ and~$Z$
  (system size $N=4000$, trails start from the lower-left corner).}
 \label{tab:FittedValues}
\end{table}

Interestingly, for trails that start in the lower-left corner of an
\emph{equilateral} triangle the fitted values of $\alpha$ and~$\beta$
\emph{do} agree for the exit distribution at different system sizes (see
table~\ref{tab:Equilateral}). In other words, the exit distribution of the
trails is symmetric in an equilateral triangle. Note that this result is not
trivial because our boundary conditions are not symmetric between the left
side and the base of the triangle, see figure~\ref{fig:BrauerTriangle}. We
infer that our choice of boundary conditions does not destroy the isotropy of
the model, and hence does not interfere with conformal invariance in the
scaling limit.

Before we consider the convergence to a limit in more detail, let us also
consider the joint distributions of $X$, $Y$ and~$Z$. Since the fits of
the marginal distributions give different values of $\alpha$ and~$\beta$
at finite system sizes, it is to be expected that the joint distributions
measured in the simulations will not be well described by those of an RBM.
This is indeed what we see when we try to fit the data for the joint
distributions to the distribution function for the RBMs, see for instance
table~\ref{tab:Equilateral}. We note that the value of~$\chi^2$ for the
fits of the joint distributions is several times the number of bins at small
system sizes. However, $\chi^2$ does go down as the system size increases,
which is a sign that the fits become better for larger systems.

If the hull of the self-avoiding trail does converge to that of a reflected
Brownian motion in the scaling limit, then the fitted values of $\alpha$
and~$\beta$ should converge to a limit value as $N\to\infty$. From the change
in the fitted values of $\alpha$ and~$\beta$ with system size observed in the
simulations we can see that if there is convergence, then it is very slow. We
make the educated guess that $\alpha$ and~$\beta$ converge with the system
size~$N$ as~$1/\log_{10} N$ (corrections that behave as a power of $\log_{10}
N$ rather than as a power of~$N$ itself are consistent with earlier
findings~\cite{martins:1998} and with the presence of a zero conformal
weight).

\begin{table}
 \centering\small
 \begin{tabular}{@{}l|cccc|cccc@{}}
   & \multicolumn{4}{c|}{Marginal distribution of~$X$}
   & \multicolumn{4}{c}{Joint distribution of $X$, $Y$ and~$Z$} \\
  $N$
   & $\alpha/\pi$ & $\beta/\pi$ & $\chi^2$ & \#bins
   & $\alpha/\pi$ & $\beta/\pi$ & $\chi^2$ & \#bins \\
  \hline
  \vrule height2.5ex width0pt
  100
   & 0.421(5) & 0.439(5) & 7754 & 99
   & 0.428(4) & 0.434(3) & 8401 & 1193 \\
  200
   & 0.440(2) & 0.449(2) & 556 & 100
   & 0.436(4) & 0.441(3) & 8577 & 1317\\
  300
   & 0.445(3) & 0.450(3) & 708 & 100
   & 0.440(4) & 0.443(4) & 8375 & 1207 \\
  400
   & 0.448(2) & 0.452(2) & 311 & 100
   & 0.442(4) & 0.444(5) & 8386 & 1193 \\
  600
   & 0.453(2) & 0.455(2) & 238 & 100
   & 0.445(4) & 0.447(4) & 8181 & 1200 \\
  800
   & 0.454(2) & 0.457(2) & 201 & 100
   & 0.448(4) & 0.450(3) & 8280 & 1310 \\
  1200
   & 0.457(2) & 0.458(2) & 148 & 100
   & 0.451(4) & 0.452(4) & 7661 & 1427 \\
  1600
   & 0.460(2) & 0.461(2) & 160 & 100
   & 0.453(4) & 0.454(4) & 7938 & 1441 \\
  2000
   & 0.460(1) & 0.461(2) & 123 & 100
   & 0.453(4) & 0.454(4) & 7392 & 1200 \\
 2400
   & 0.461(2) & 0.462(1) & 110 & 100
   & 0.455(4) & 0.455(4) & 7102 & 1083 \\
  3200
   & 0.462(2) & 0.462(2) & 152 & 100
   & 0.455(4) & 0.456(3) & 7228 & 1317 \\
  4000
   & 0.464(2) & 0.464(2) & 126 & 100
   & 0.457(3) & 0.457(3) & 7118 & 1434
 \end{tabular}
 \caption{Fitted values for $\alpha$ and~$\beta$ together with the value
  of~$\chi^2$ and the number of bins for two different distributions of
  a trail on an equilateral triangle, started from the lower-left corner.}
 \label{tab:Equilateral}
\end{table}

To test convergence, we therefore make a linear fit of the fitted values of
$\alpha$ and~$\beta$ for the different distributions on different triangles
against $1/\log_{10} N$. We accept the linear fit if it passes each of the
following three tests:
\begin{enumerate}
\item We look at the value of~$\chi^2$ for the fit. If this value exceeds
 the $10\%$ probability threshold for the $\chi^2$ distribution, we reject
 the fit.
\item We compare the value of the intercept at $N=\infty$ from the linear
 fit with the value we obtain if we fit the data to a parabola in
 $1/\log_{10} N$. If the values do not agree within 1.96~standard deviation
 (the Gaussian $5\%$ level) of the linear fit, the fit is rejected.
\item We do a run test. If the data is well described by a line, then each
 data point should be independently above or below this line with equal
 probabilities. Thus we can predict how many consecutive runs of points above
 the line and below the line to expect. If the probability for the number of
 runs we find is less than~$12\%$, the fit is rejected.
\end{enumerate}
For more background on these (and other) kinds of tests for data fitting,
the reader can consult for instance Barlow~\cite{barlow:1989}.

We have a total of 224 sets of data points for which we attempt a linear fit
(3 marginal distributions plus 4 joint distributions for the variables $X$,
$Y$ and~$Z$, times 8 different base angles, times 2 because the trails can
start either from the top or the lower-left corner, times 2 variables $\alpha$
and $\beta$). Of these, 39 (17\%) give an accepted linear fit in the 12~system
sizes, and an additional 27 (for a total of 66, i.e.~29\%) give an accepted
linear fit if we leave out the smallest system size. For the other sets of
data we can not observe the convergence from our simulations without further
knowledge of how $\alpha$ and~$\beta$ should behave as functions of~$N$.

\begin{figure}[t!]
\includegraphics{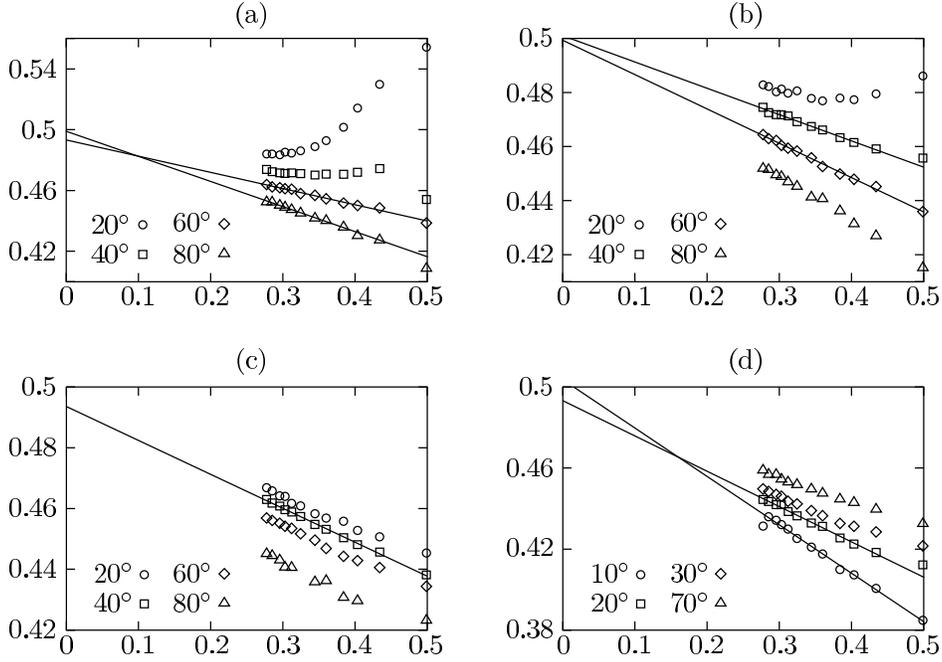}
 \caption{Graphs (a), (b) and~(c) show the fitted values of $\beta/\pi$ against
  $1/\log_{10} N$ for the marginal distributions of~$X$ and~$Z$ and the joint
  distribution of $X$, $Y$ and~$Z$, respectively, for trails started from
  the left corner. Graph (d) shows the fitted values of $\alpha/\pi$ against
  $1/\log_{10} N$ for the joint distribution of $X$, $Y$ and~$Z$ for trails
  started from the top. In those cases where a linear fit is accepted, the
  fitted line is shown as well.}
 \label{fig:LinearFits}
\end{figure}

From each accepted linear fit of $\alpha$ or~$\beta$ against $1/\log_{10} N$
we obtain a value for the intercept with the $\alpha$- or~$\beta$-axis as
$N\to\infty$. Considering only the accepted linear fits with all 12 system
sizes taken into account, these intercept values give $\alpha/\pi\to
0.4964(48)$ and $\beta/\pi\to0.4952(75)$ as $N\to\infty$. If we include
the additional accepted linear fits where the smallest system size is left
out, we obtain $\alpha/\pi\to0.4955(54)$ and $\beta/\pi\to0.4952(66)$. These
results are consistent with the hypothesis that the hull of the self-avoiding
trail in the scaling limit is the same as that of an RBM$_{\pi/2,\pi/2}$, that
is, a reflected Brownian motion with perpendicular reflection on the boundary.

\section{Percolation and the last-visit distribution}
\label{sec:lastvisit}

In this section we look at a different distribution for the self-avoiding
trail of the Brauer model. The motivation for this comes from a further
connection between the percolation exploration process and RBM$_{\pi/3,\pi/3}$
discovered by Dub\'edat~\cite{dubedat:2004c}, the analogue of which we want
to investigate for the Brauer model. But first let us consider the case of
percolation. In section~\ref{sec:brauer} we explained that the Brauer model
studied in this paper becomes a model for critical bond percolation if we
take the probabilities of the vertex states $a$ and~$b$ on
page~\pageref{vertexstates} proportional to $\sin(2\phi/3)$ and
$\sin((\pi-2\phi)/3)$, respectively, and $p(c)=0$. The paths defined by
the model then correspond to the boundaries of the percolation clusters, or
equivalently, to a percolation exploration process. The hull generated by a
path of the model should therefore have the same distribution as the hull of
an RBM$_{\pi/3,\pi/3}$ in the scaling limit, as we discussed in
section~\ref{sec:locality}.

Similar simulations as for the self-avoiding trails allow us to test this
hypothesis. We have measured in the same way as before the joint distribution
of the variables $X$, $Y$ and~$Z$ on different triangles with fixed system
size $N=2000$. Our hypothesis is that this joint distribution is predicted
by equation~(\ref{equ:joint}) for reflected Brownian motion with reflection
angles $\alpha$ and~$\beta$, where $\alpha$ and~$\beta$ should be equal to
$\pi/3$. Table~\ref{tab:Percolation} shows the results of a least-squares
fit for the reflection angles $\alpha$ and~$\beta$, which agree with the
hypothesis that the hull distribution is the same as that of an RBM$_{\pi/3,
\pi/3}$.

\begin{table}
 \centering\small
 \begin{tabular}
  {@{}l|c|cc|cc@{}}
   & & \multicolumn{2}{c|}{from the top}
     & \multicolumn{2}{c@{}}{from the left corner} \\
   & angle &
   $\alpha/\pi$ & $\beta/\pi$ & $\alpha/\pi$ & $\beta/\pi$ \\
  \hline
   \vrule height2ex width0pt
    & $20^\circ$ & 0.331(4) & 0.350(4) & 0.334(2) & 0.334(2) \\
   joint
    & $40^\circ$ & 0.334(4) & 0.339(3) & 0.333(1) & 0.334(2) \\
   distribution
	& $60^\circ$ & 0.334(2) & 0.333(1) & 0.333(2) & 0.335(2) \\
	& $80^\circ$ & 0.333(1) & 0.331(1) & 0.331(2) & 0.335(2) \\
  \hline
   \vrule height2ex width0pt
    & $20^\circ$ & 0.336(2) & 0.335(2) & 0.332(2) & 0.329(3) \\
   last-visit
    & $40^\circ$ & 0.329(4) & 0.331(3) & 0.332(3) & 0.332(2) \\
   distribution
	& $60^\circ$ & 0.332(3) & 0.334(3) & 0.332(3) & 0.332(3) \\
	& $80^\circ$ & 0.328(4) & 0.339(3) & 0.327(3) & 0.333(3)
 \end{tabular}
 \caption{Fitted values for $\alpha$ and~$\beta$ on different triangles
  for the joint and last-visit distributions of the exploration process
  of critical bond percolation at system size $N=2000$.}
 \label{tab:Percolation}
\end{table}

It was shown by Dub\'edat~\cite{dubedat:2004c} that there exists a rather
surprising further connection between reflected Brownian motion and the
exploration process of critical percolation. To explain this connection,
suppose that~$W$ is a local process in the triangle $T=T_{\lambda,\mu}$
started from $w=w_{\lambda,\mu}$, where $\lambda,\mu$ are fixed angles
such that $\lambda+\mu<\pi$. Let $\tau:=\inf\{t\geq0:W(t)\in[0,1]\}$, and
let~$\sigma$ be the last time before~$\tau$ when~$W$ visits the boundary
of~$T$. Let $E$ denote the event that $W(\sigma)$ is on the right side of
the triangle. Then we can consider the probability of the event~$E$
conditioned on~$\{W(\tau)=x\}$. We call this conditional probability, which
is a function of~$x$, the \emph{last-visit distribution} of~$W$ in~$T$.

For reflected Brownian motion this last-visit distribution can be computed
using the fact that the exit distribution is uniform in a well-chosen
triangle. For the case of an RBM$_{\pi/3,\pi/3}$ in an equilateral triangle,
this computation was done by Dub\'edat in~\cite{dubedat:2004a}. The
generalization to an RBM$_{\alpha,\beta}$ in the triangle~$T_{\lambda,\mu}$
started from~$w_{\lambda,\mu}$ is straightforward, and gives
\begin{equation}
 \Prob[E\mid W(\tau)=x]
 = F_{\pi-\alpha,\pi-\beta}\left( F^{-1}_{\lambda,\mu}(x) \right),
 \label{equ:lastvisit}
\end{equation}
where~$F_{\gamma,\delta}$, given by equation~(\ref{equ:trianglemapping}),
is the unique conformal transformation of the upper half-plane onto
$T_{\gamma,\delta}$ which fixes $0$ and~$1$ and sends~$\infty$ to
$w_{\gamma,\delta}$.

The last-visit distribution can also be computed rigorously for chordal
SLE(6), as was shown by Dub\'edat~\cite{dubedat:2004c}. The obtained formula
is exactly the formula~(\ref{equ:lastvisit}) in the case of an RBM$_{\pi/3,
\pi/3}$ (i.e.\ $\alpha=\beta=\pi/3$). This fact is rather surprising if one
considers the very different ways in which the two processes explore space:
whereas an SLE(6) process never crosses itself, an RBM$_{\pi/3,\pi/3}$
crosses itself many times. In particular, it is known that on the event~$E$,
the last point visited on the right side of the triangle~$T$ by an SLE(6)
process must be also the lowest point of the hull on the right side. For
an RBM$_{\pi/3,\pi/3}$, however, the last point visited on the right side
is almost surely not the lowest point of the hull.

\begin{figure}
\includegraphics{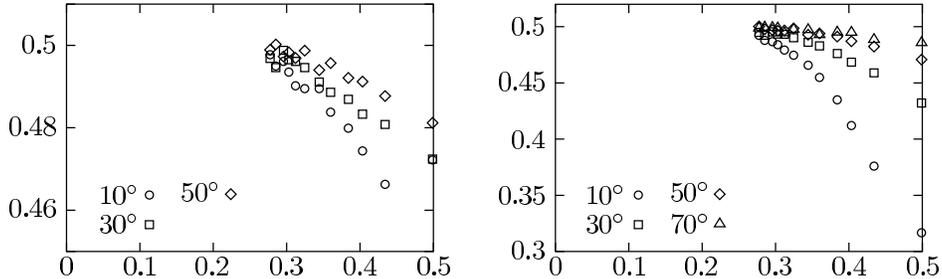}
\caption{Fitted values of $\alpha/\pi$ (left) and $\beta/\pi$ (right)
 plotted against $1/\log_{10} N$ for the last-visit distribution of trails
 starting from the lower-left corner of different triangles.}
\label{fig:LastVisitFits}
\end{figure}

Let us now return to our simulations of the exploration path of critical
percolation. It is quite easy to count in these simulations the number of
paths that land in a given bin and visited the side to the left (and not to
the right) of the starting point last before reaching the opposite side.
Divided by the total number of paths that land in a given bin, this binned
data can be compared to the last-visit distribution of
equation~(\ref{equ:lastvisit}). Note, however, that this binned data
represents the conditional probability of~$E$ given that the path lands in
an interval rather than at a given point~$x$. To make the comparison we
therefore have to integrate~(\ref{equ:lastvisit}) with respect to the exit
distribution of the exploration path.

We take the exit distribution for the paths in the simulations to be well
described by the least-squares fit for the marginal distribution of~$X$.
Then we can integrate~(\ref{equ:lastvisit}) numerically (with $\alpha$ and
$\beta$ as parameters) with respect to this fixed exit distribution, and
obtain from this the probability of~$E$ conditioned on the event that the
path lands in a given bin. This can be used to make a least-squares fit
for the parameters $\alpha$ and~$\beta$ of the last-visit distribution.
Table~\ref{tab:Percolation} shows that our simulations for critical
percolation are in agreement with the fact that the last-visit distribution
is given by that of an RBM$_{\pi/3,\pi/3}$.

A similar analysis can be carried out for the self-avoiding trails of the
Brauer model. As before, here we face the problem that the model converges
only slowly. We therefore make least-squares fits of $\alpha$ and~$\beta$ for
different system sizes, and plot the results against $1/\log_{10} N$, see
figure~\ref{fig:LastVisitFits}. It is clear that here we can not conclude
what the scaling limit is, without an analytic prediction of how the
parameters $\alpha$ and~$\beta$ behave as functions of~$N$. However, our
results do not exclude the possibility that the last-visit distribution for
the self-avoiding trails in the scaling limit is the same as that of an
RBM$_{\pi/2,\pi/2}$, just like the distribution of the hull.

\paragraph{Conclusions}

We have carried out a classical Monte Carlo study of the hull distribution
for the Brauer model at $q=1$. Our results are in agreement with the
hypothesis that the hull distribution coincides with that of a reflected
Brownian motion with perpendicular reflection. We have tested our methods
for percolation, and found good agreement with the fact that in this case
the hull must have the same distribution as that of a Brownian motion which
is reflected at angles of $60^\circ$ with respect to the boundary. We have
also studied the last-visit distribution, but for the Brauer model our
simulation results are inconclusive.

\paragraph{Acknowledgements}

We thank Wendelin Werner for useful discussions in the early stages of this
project. This work is part of the research programme of the `Stichting voor
Fundamenteel Onderzoek der Materie (FOM)', which is financially supported by
the `Nederlandse Organisatie voor Wetenschappelijk Onderzoek (NWO)'.


\begin{thebibliography}{10}
\providecommand{\eprint}[2][]{\url{[#2]}}

\bibitem{ahlfors:1966}
Ahlfors, L.\,V. (1966, 2nd edition).
\newblock {\em Complex analysis: an introduction to the theory of analytic
  functions of one complex variable\/}.
\newblock New York: McGraw-Hill.

\bibitem{barlow:1989}
Barlow, R.\,J. (1989).
\newblock {\em Statistics. A guide to the use of statistical methods in the
  physical sciences\/}.
\newblock Chichester: John Wiley \& Sons.

\bibitem{baxter:1978}
Baxter, R.\,J. (1978).
\newblock Solvable eight-vertex model on an arbitrary planar lattice.
\newblock {\em Philos.\ Trans.\ Roy.\ Soc.\ London Ser.\ A\/} {\em 289\/}, pp.
  315--346.

\bibitem{baxter:1982}
Baxter, R.\,J. (1982).
\newblock {\em Exactly solved models in statistical mechanics\/}.
\newblock London: Academic Press.

\bibitem{camia:2005}
Camia, F. \& Newman, C.\,M. (preprint).
\newblock The full scaling limit of two-dimensional critical percolation.
\newblock \eprint{math.PR/0504036}.

\bibitem{dubedat:2004a}
Dub\'edat, J. (2004).
\newblock Reflected planar {B}rownian motions, intertwining relations and
  crossing probabilities.
\newblock {\em Ann.\ Inst.\ H.\ {Poincar\'e} Probab.\ Statist.\/} {\em 40\/},
  pp. 539--552.
\newblock \eprint{math.PR/0302250}.

\bibitem{dubedat:2004c}
Dub\'edat, J. (preprint).
\newblock Excursion decompositions for {SLE} and {W}atts' crossing formula.
\newblock \eprint{math.PR/0405074}.

\bibitem{degier:2005}
Gier, J.~de \& Nienhuis, B. (2005).
\newblock Brauer loops and the commuting variety.
\newblock {\em J.\ Stat.\ Mech.\ Theory Exp.\/} 006, 10 pp.

\bibitem{gunn:1985}
Gunn, J.\,M.\,F. \& Ortu{\~n}o, M. (1985).
\newblock Percolation and motion in a simple random environment.
\newblock {\em J.\ Phys.\ A\/} {\em 18\/}, pp. L1095--L1101.

\bibitem{kager:2004}
Kager, W. (preprint).
\newblock Reflected Brownian motion in generic triangles and wedges.
\newblock Submitted to {\em Stoch.\ Process.\ Appl.}
\newblock \eprint{math.PR/0410007}.

\bibitem{kenyon:2003}
Kenyon, R. (preprint).
\newblock An introduction to the dimer model.
\newblock \eprint{math.CO/0310326}.

\bibitem{kim:1987}
Kim, D. \& Pearce, P.\,A. (1987).
\newblock Scaling dimensions and conformal anomaly in anisotropic lattice
spin models.
\newblock {\em J.\ Phys.\ A\/} {\em 20}, pp. L451--L456.

\bibitem{kong:1989}
Kong, X.\,P. \& Cohen, E.\,G.\,D. (1989).
\newblock Anomalous diffusion in a lattice-gas wind-tree model.
\newblock {\em Phys.\ Rev.\ B\/} {\em 40\/}, pp. 4838--4845.

\bibitem{lsw:2001a}
Lawler, G.\,F., Schramm, O. \& Werner, W. (2001).
\newblock Values of {Brownian} intersection exponents {I}: {Half-plane}
  exponents.
\newblock {\em Acta Math.\/} {\em 187\/}, pp. 237--273.
\newblock \eprint{math.PR/9911084}.

\bibitem{lsw:2003}
Lawler, G.\,F., Schramm, O. \& Werner, W. (2003).
\newblock Conformal restriction: the chordal case.
\newblock {\em J.\ Amer.\ Math.\ Soc.\/} {\em 16\/}, pp. 917--955.
\newblock \eprint{math.PR/0209343}.

\bibitem{lyklema:1985}
Lyklema, J.\,W. (1985).
\newblock The growing self-avoiding trail.
\newblock {\em J.\ Phys.\ A\/} pp. L617--L624.

\bibitem{martins:1998}
Martins, M.\,J., Nienhuis, B. \& Rietman, R. (1998).
\newblock Intersecting loop model as a solvable super spin chain.
\newblock {\em Phys.\ Rev.\ Lett.\/} {\em 81\/}, pp. 504--507.

\bibitem{reshetikhin:1983}
Reshetikhin, N.\,Y. (1983).
\newblock The functional equation method in the theory of exactly soluble
  quantum systems.
\newblock {\em Sovjet Phys.\ JETP\/} {\em 57\/}, pp. 691--696.

\bibitem{ruijgrok:1988}
Ruijgrok, T.\,W. \& Cohen, E.\,G.\,D. (1988).
\newblock Deterministic lattice gas models.
\newblock {\em Phys.\ Lett.\ A\/} {\em 133\/}, pp. 415--418.

\bibitem{schultz:1981}
Schultz, C.\,L. (1981).
\newblock Solvable $q$-state models in lattice statistics and {Q}uantum {F}ield
  {T}heory.
\newblock {\em Phys.\ Rev.\ Lett.\/} {\em 46\/}, pp. 629--632.

\bibitem{smirnov:2001}
Smirnov, S. (2001).
\newblock Critical percolation in the plane: conformal invariance, {Cardy's}
  formula, scaling limits.
\newblock {\em C.\ R.\ Acad.\ Sci.\ Paris S\'er.\ I Math.\/} {\em 333\/}, pp.
  239--244.

\bibitem{varadhan:1985}
Varadhan, S.\,R.\,S. \& Williams, R.\,J. (1985).
\newblock Brownian motion in a wedge with oblique reflection.
\newblock {\em Comm.\ Pure App.\ Math.\/} {\em 38\/}, pp. 405--443.

\bibitem{devega:1987}
Vega, H.\,J.~de \& Karowski, M. (1987).
\newblock Exact {B}ethe {A}nsatz solution of {O($2n$)} symmetric theories.
\newblock {\em Nucl.\ Phys.\ B\/} {\em 280\/}, pp. 225--254.

\bibitem{werner:2001}
W. Werner (2001).
\newblock Critical exponents, conformal invariance and planar Brownian motion.
\newblock In: C. Casacuberta, R.M. Miro-Roig, J. Verdera and
S. Xambo-Descamps, eds., {\em European Congress of Mathematics, Vol.~II
(Barcelona, 2000)}, pp.  87--103, {\em Progr.\ Math.} {\bf 202}.
Basel: Birkh\"auser,
\newblock \eprint{arXiv:math.PR/0007042}.

\bibitem{werner:2004b}
Werner, W. (2004).
\newblock Random planar curves and {S}chramm-{L}oewner evolutions.
\newblock In {\em Lectures on probability theory and statistics\/}, vol. 1840
  of {\em Lecture notes in Math.\/}, pp. 107--195. Berlin: Springer.
\newblock \eprint{math.PR/0303354}.

\bibitem{ziff:1991}
Ziff, R.\,M., Kong, X.\,P. \& Cohen, E.\,G.\,D. (1991).
\newblock Lorentz lattice-gas and kinetic-walk model.
\newblock {\em Phys.\ Rev.\ A\/} {\em 44\/}, pp. 2410--2428.

\end{thebibliography}
\end{document}